\documentclass[aps,twocolumn,showpacs]{revtex4}
\usepackage{graphics}
\usepackage{latexsym}
\begin{document}
\bibliographystyle{apsrev}
\def\half{{1\over 2}}
\def \D {\mbox{D}}
\def\curl {\mbox{curl}\,}
\def \ep {\varepsilon}
\def \lleq {\lower0.9ex\hbox{ $\buildrel < \over \sim$} ~}
\def \ggeq {\lower0.9ex\hbox{ $\buildrel > \over \sim$} ~}
\def\beq{\begin{equation}}
\def\eeq{\end{equation}}
\def\ber{\begin{eqnarray}}
\def\eer{\end{eqnarray}}
\def \apl {ApJ, }
\def \aps {ApJS, }
\def \pd {Phys. Rev. D, }
\def \prl {Phys. Rev. Lett., }
\def \pl {Phys. Lett., }
\def \np {Nucl. Phys., }
\def \l {\Lambda}


\title   {A VIABLE COSMOLOGY WITH A SCALAR FIELD COUPLED TO THE TRACE OF THE STRESS-TENSOR} 

\author{M.Sami}
\altaffiliation[On leave from:]{ Department of Physics, Jamia Millia, New Delhi-110025}
\email{sami@iucaa.ernet.in}
\author{T. Padmanabhan}
\email[]{nabhan@iucaa.ernet.in }
\homepage[]{http://www.iucaa.ernet.in/~paddy}
\affiliation{IUCAA, Post Bag 4, Ganeshkhind,\\
 Pune 411 007, India.}

\begin{abstract}

We study the cosmological evolution of a scalar field that couples to the trace
  $T=T^{a}_a $ 
of energy momentum tensor of all the fields (including
itself). In the case of a  shallow exponential potential, the presence
of  coupling to the trace $T$  in the field equation   makes the energy density of the scalar field decrease faster thereby hastening the 
commencement of radiation domination. This effect gradually diminishes at later epochs allowing the scalar field to  dominate the energy density again. We interpret this phase as the current epoch of
 cosmic acceleration with $\Omega_{\phi}=0.7$. A variant of this model can lead to accelerated expansion at the present epoch
followed by a $a(t)\propto t^{2/3}$ behaviour as $t\to \infty$, 
 making the model free 
from future event horizon. The main features of the model are  independent of initial conditions. However, 
fine tuning of parameters is necessary for viable evolution.
\end{abstract}

 \maketitle
 \section {INTRODUCTION}
The interpretation of current observations in terms of an accelerating universe requires an exotic form of energy density with $(\rho+3p)<0$. A wide variety of scalar field models have been conjectured for this purpose
including quintessence \cite{phiindustry}, K-essence\cite{armen}, tachyonic scalar fields\cite{tachyonindustry,tptachyon,bjp} with the last one being originally motivated by string theoretical ideas \cite{ashoke}. (For a recent review of issues
related to the cosmological constant and dark energy, see \cite{tpcc}). Since all these models have a free, undetermined function $V(\phi)$, it is possible to incorporate any reasonable $a(t)$ in any of these models \cite{tptachyon}. Thus, at a fundamental level, all these models can be objected to --- quite correctly --- as lacking in predictive power. The appropriate way of judging these models, therefore, will be based on the
physical input that is used in constructing the model and the generic features which arise in the model. 

We investigate here a model in which the coupling of the scalar field to the matter is based on a physical principle which can be motivated as follows:
The most obvious generalization of the Newtonian gravity, based on $\nabla^2\phi=4\pi G\rho$
will be $\Box \phi=4\pi G T$ in which  the trace $T=T^a_a$ of the energy momentum of fields act as a source for
a scalar field.  This theory, however, has to be nonlinear, since the scalar field itself has a nonzero trace for the energy momentum tensor which should appear on the right hand side. Further, since cosmological constant has a nonzero trace for stress tensor, this scalar field will also couple to the cosmological constant and the effective cosmological constant will become dynamical. Such a model was developed and investigated more than a decade back \cite{jvntp,tpstp88}, as a possible solution to cosmological constant problem. This work, however, was ambitious in the sense that it attempted to tackle the problem {\it without any fine tuning} and did not succeed \cite{tpstp88}. The current thinking in cosmology is remarkably tolerant to fine tuning of parameters and virtually every scalar field model suggested in the literature has either explicit or implicit fine tuning of parameters. In view of this, one can resurrect the above model, in which the form of the interaction arises from a clear physical requirement, and explore its consequences for cosmology. We attempt to do this in this paper and show the the resulting models have very nice properties.
 
\section{DYNAMICS OF THE SCALAR FIELD WITH A COUPLING TO $T^a_a$}

The Lagrangian for a system made of the gravitational field, matter sources and the scalar field which couples to the trace of the total energy momentum tensor is derived in \cite{tpstp88}. (This derivation is briefly summarized in Appendix A for completeness.)
 The complete action, which
 takes into account the trace coupling of the scalar field, has the form (see Appendix A):
\begin{eqnarray}
S &=& \nonumber (16 \pi G)^{-1} \int{R \sqrt{-g}} d^4x+{1 \over 2} \int { \beta(\phi) \phi^i \phi_i \sqrt{-g}}d^4x \\ 
&-&  (8 \pi G)^{-1} \int{\alpha(\phi) V(\phi)\sqrt{-g}}d^4x + S_{\rm source}
 \label{action}
\end{eqnarray}
where 
the coupling to the trace is given by an arbitrary  function $f(\phi)$ and
\begin{equation}
 \alpha(\phi)={1 \over {1+4f(\phi)}}, \quad \beta(\phi)={1 \over {1+2f(\phi)}}
 \end{equation}
 The $S_{\rm source}$ is the remaining source term (radiation, matter ... ), with appropriate coupling
 to $\phi$ (see Appendix A).  
The energy momentum tensor for the field $\phi$  which arises from the action (\ref{action})
 is given by
\begin{equation}
T^{ik}=\alpha(\phi)V(\phi)g^{ik}+\beta(\phi) \left[\phi^i \phi^k-{1 \over 2}g^{ik} \phi^j \phi_j \right]
\label{tikeqn}
\end{equation}
 We shall assume that scalar field is evolving in an isotropic and homogeneous space time and that $\phi$ is a function
of time alone.
The energy density $\rho_{\phi}$ and  pressure $p_{\phi}$ 
 obtained from $T^{ik}$ are
\begin{equation}
\rho_{\phi}=\beta(\phi){ \dot{\phi}^2 \over 2}+\alpha(\phi)V(\phi),~~~~p_{\phi}=\beta(\phi){ \dot{\phi}^2 \over 2}-\alpha(\phi)V(\phi)
\end{equation}
The coupling to $T^a_a$  modifies the field energy density $\rho_{\phi}$, pressure $p_{\phi}$ and the equation
of state parameter $w_{\phi}=(p_{\phi}/\rho_{\phi})$.

In a universe containing dust like matter and radiation, 
the action (\ref{action}) leads to
 following  evolution equations for the field:
\begin{eqnarray}
\ddot{\phi}+{3\dot{a} \over a}\dot{\phi}&=& \nonumber{ \dot{\phi}}^2 {f_{,\phi} \over {1+2f}} 
+4V(\phi) {{(1+2f)} \over {(1+4f)}^2}f_{,\phi}\\
&-&V_{,\phi}(\phi)
{{(1+2f)} \over (1+4f)}+{\rho^i_m \over a^3}f_{,\phi}{ (1+2f) \over {(1+f)^2}}
\label{evoleqn}
\end{eqnarray}
(The radiation, being traceless, does not couple to $\phi$.)
  The  Friedman equation with the modified
energy momentum tensor given by (\ref{tikeqn}) is
\begin{equation}
{\dot{a}^2(t) \over a^2(t)}={{8 \pi G} \over 3}\left[{\dot{\phi}^2 \over {2(1+2f)}} +{ V(\phi) \over {1+4f}}+{ \rho_b(a) }
\right]
\label{friedeqn}
\end{equation}
where  background energy density due to radiation and matter is given by

\begin{equation}
\rho_b(a)={ \rho^i_R \over a^4}+{\rho^i_m \over {a^3(1+f)}}
\end{equation}
The simplest form of coupling which we shall adopt is   the linear one
\begin{equation}
 f(\phi)=g_c( \phi/M_p) 
 \end{equation}
where $  M_p =\sqrt{1 / (8 \pi G)}$ is the reduced Planck's mass and $g_c$ is a coupling constant.
Since  equations (\ref{evoleqn}) and (\ref{friedeqn}) reduces to the standard form 
for  $g_c=0$, this parameter  gives the strength of the forcing term. 
Equation (\ref{evoleqn}) describes the evolution of the scalar field in the expanding universe under the influence of an external ``force'', which depends
upon the field and its kinetic energy $(1/2)\dot{\phi}^2$. The influence of this term on the dynamics of the scalar field is accumulative.

In the absence of matter, the conservation equation formally equivalent to (\ref{evoleqn}), has the usual form
\begin{equation}
\dot{\rho_{\phi}}+3H(\rho_{\phi}+p_{\phi})=0
\label{conseqn}
\end{equation}
and the evolution of energy density is given by
\begin{equation}
\rho_{\phi}=\rho_{0\phi}e^{-\int{6\left (1-\zeta(a)\right){da \over a}}}
\end{equation}
with
$$ \zeta(a)={1 \over {(K_e/ P_e) +1}}$$
where the ratio of effective kinetic to potential energy$ (K_e/P_e)$ is given by
\begin{equation}
{K_e \over P_e}={\beta(\phi) \over \alpha(\phi)}{\dot{\phi}^2 \over {2 V(\phi)}}
\end{equation}
Interaction modifies the kinetic as well as the potential energy through $ \alpha(\phi)$ and $\beta(\phi)$, however, $ \rho_{\phi}$ bears the same relation to their ratio as in the case of standard scalar field cosmology. 
Equation  (\ref{conseqn}), of course,  is not valid in presence of matter; neverthless numerical work shows that
 the ratio $(K_e / P_e)$ is still  a good parameter which influences the dynamics. 
The evolution of potential to kinetic energy ratio plays a significant role in  the growth or decay of the energy density $\rho_\phi$ 
at a given epoch and will be crucial in the following discussion.

 Let us next address the question of choice of potential of the scalar field which would lead to a viable cosmological
 model. The obvious restriction on the evolution  is that,  
 starting  from Planck's time,  the scalar field should survive till
today (to account for the observed late time accelerated expansion)
 without interfering with the nucleosynthesis of the standard model.

Exponential potentials  lead to solutions which are the attractors of evolution equations 
 when $g_c=0$ and provide backdrop for the understanding of dynamics in our case. A standard 
model with   
\begin{equation}
V(\phi)=V_0 \exp\left[-\lambda{ \phi \over M_p}\right]
\label{exppot}
\end{equation}
(and $g_c=0$) is well studied in literature and 
can be divided into two categories: (a) The exponential potentials for which $\rho_{\phi}$ scales slower than 
the background density $\rho_b=(1/a^n)$ which translates to $\lambda< \sqrt n$. (b) Potentials for which scalar field energy density scales faster than the $\rho_b$, i.e.
$\lambda>\sqrt n$. In the first case 
if the background energy density was sub-dominant, it would become more sub-dominant and radiation domination 
will never occur.
In the second case there exists a scaling solution which  mimics the background energy density that is dominant, with
\begin{equation}
\Omega_{\phi}={\rho_{\phi} \over {\rho_{\phi}+\rho_b}}={n \over \lambda^2}; \quad \rho_{\phi}\propto
 \frac{1}{a^n}
\end{equation}

The discussion of this model with $g_c\ne 0$ (so that the coupling to the trace of the stress tensor is switched on) is given in Appendix B. It turns out that this model is not as attractive as another variant of the potential with $V(\phi)=V_0 e^{\lambda{ \phi \over M_p}}$ which we shall concentrate on next.


\begin{figure}[p]
\resizebox{3.0in}{!}{\includegraphics{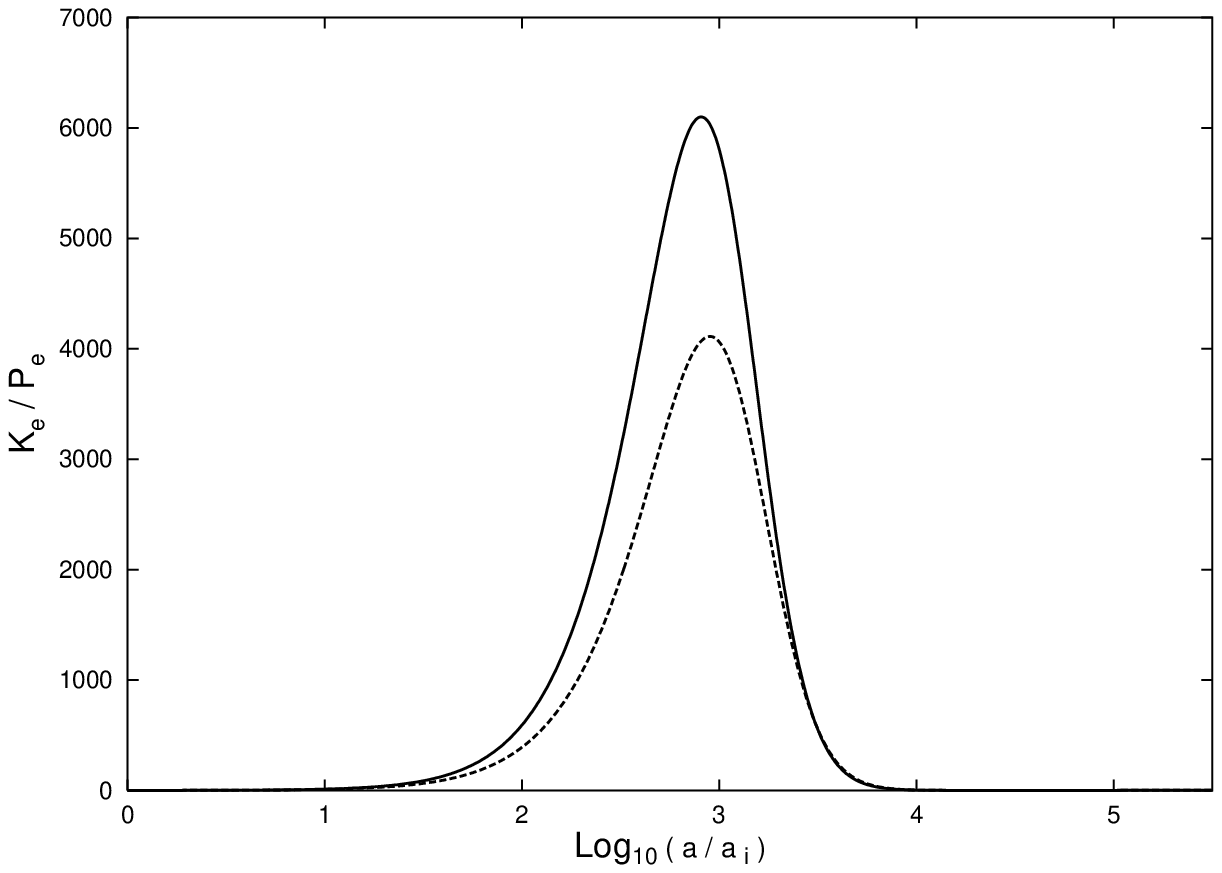}}
\caption{
The ratio of  kinetic  to potential energy of scalar field is depicted for the exponential potential (\ref{pottw+} ) with
$\lambda=5.2$ , $V_0^{1/4}\simeq 2.8\times 10^{-30} M_p$ with two different values for coupling
(chosen for illustration): $g_c=0.04$
(dashed line) and $ g_c=0.041$(solid line).The kinetic energy ( as compared to the potential energy ) is seen to build up fast to
a large value forcing $\rho_{\phi}$ into the kinetic regime which otherwise would scale slowly with the scale factor 
($\rho_{\phi} \propto a^{-\lambda^2}$) for the model with the potential in (\ref{pottw+}). 
}
\label{figkp+}
\end{figure}

\begin{figure}[p]
\resizebox{3.0in}{!}{\includegraphics{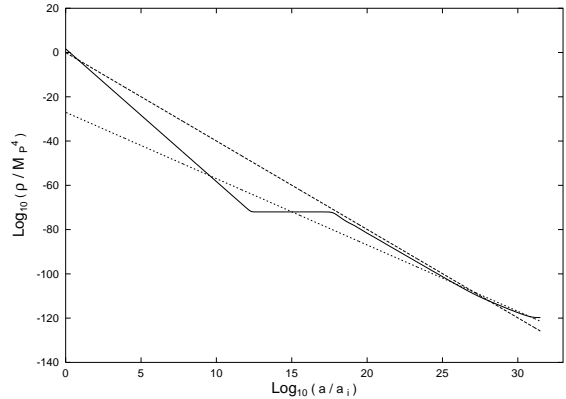}}
\caption{The energy density is plotted against the scale factor:  solid  line corresponds to $\rho_{\phi}$ for $g_c=5.9$ in case
of the model described by  (\ref{pottw+}) with $\lambda =0.52$ and $V_0^{1/4}\simeq 2.8\times 10^{-30} M_p$. The dashed and dotted lines corresponds to energy densities of matter and radiation respectively.
The  large value of the scalar field kinetic energy (relative to the potential energy ) induced by the forcing term in the
evolution
equation  ensures that the $\rho_\phi$ drops below $\rho_{b}$, after which  $\rho_{\phi}$  remains approximately constant for a period of time. At the end of this phase, the scalar field  tracks the background energy density before becoming dominant and driving the current accelerated expansion of the universe.  }
\label{figden1}
\end{figure}

\begin{figure}[p]
\resizebox{3.0in}{!}{\includegraphics{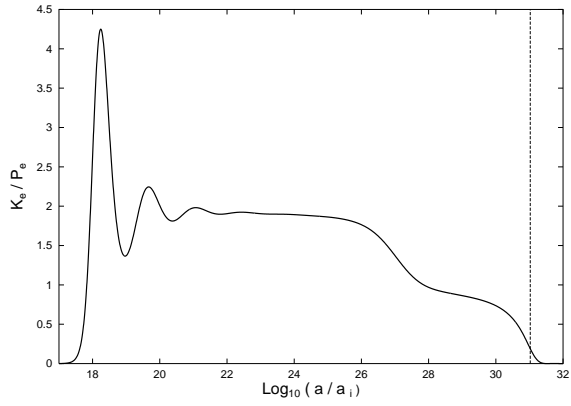}}
\caption{Evolution of the kinetic to potential energy ratio of scalar field for the model described in figure (\ref{figden1})
starting
from the locking regime onwards. At the end of locking period, the ratio fluctuates about its background value and then nearly tracks it
slowly decreasing towards its attractor value. The ratio does not lock itself to its attractor value and approaches zero where
it stays asymptotically. The vertical dashed line marks the current epoch.}
\label{figkp1+}
\end{figure}

\begin{figure}[p]
\resizebox{3.0in}{!}{\includegraphics{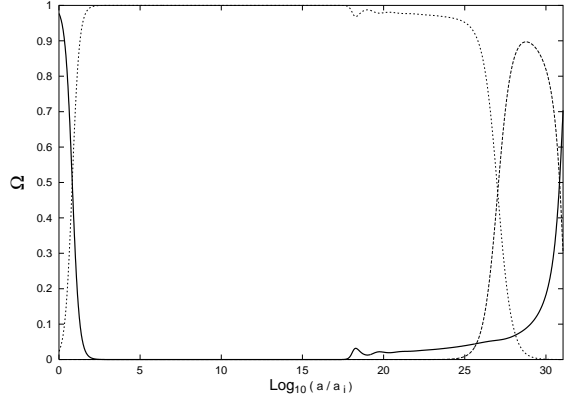}}
\caption{Evolution of dimensionless density parameter $\Omega$ is shown as a function of the scale factor  for
the model described in figure (\ref{figden1}) for: (i) the scalar field (solid line), (ii) radiation (dotted line) and
matter (dashed line). Late
time behavior of  the scalar field leads to  the present day value of $\Omega_{\phi}=0.7$ and $\Omega_{m}=0.3$.}
\label{figomega}
\end{figure}

\begin{figure}[p]
\resizebox{3.0in}{!}{\includegraphics{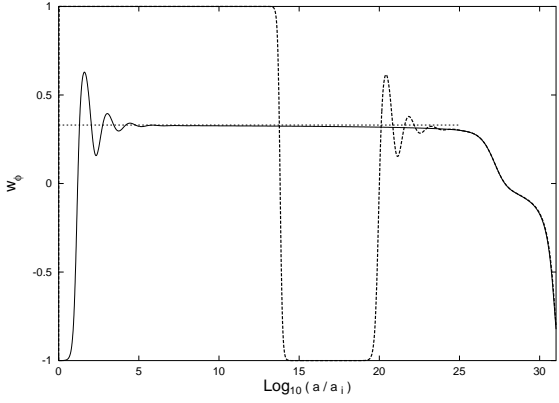}}
\caption{The  equation of state parameter of scalar field for the exponential potential (\ref{pottw+}) showing 
 different initial conditions  converge to attractor solution at late times. The horizontal
dashed line depicts the $w_b=1/3=w_R$. The solid and dashed lines start with widely different
initial conditions but lead to similar behaviour at late times.}
\label{figoustat}
\end{figure}

\begin{figure}[p]
\resizebox{3.0in}{!}{\includegraphics{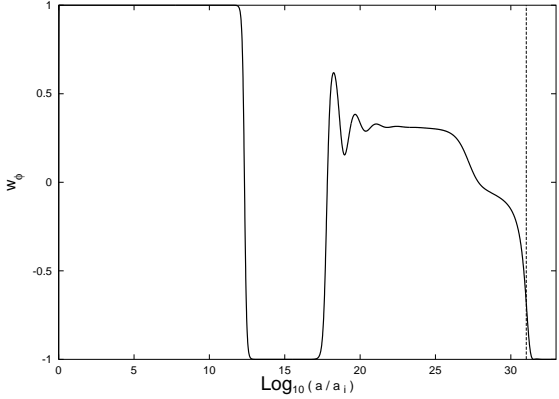}}
\caption{Evolution of equation of state parameter  of scalar field.  The coupling to the trace $T$  brings scalar field
(which was slowly rolling
down the shallow potential)
 into kinetic regime. The scalar field stays in this regime for a long time before 
 $\rho_\phi$ grows towards $\rho_{\rm b}$. During this phase, $w_\phi$
  gets locked to $w_{\phi}=-1$ ( locking regime). After the locking period
ends, $\rho_{\phi}$ tracks the background energy density slowly approaching it from below with equation of state $w_{\phi} \simeq
1/3$. At late times $\rho_{\phi}$ overtakes the background and becomes dominant to account for the current epoch of cosmic acceleration. During this time $w_{\phi}$ decreases and heads towards its inflationary attractor value ($ w_{\phi}=(\lambda^2-3)/3)$). However, before it could settle
there, the slowly moving field $\phi$ towards the origin gets trapped near the barrier formed there due to coupling to the trace $T$ . The
equation of state locks to $-1$  permanently. The vertical dashed line marks the present epoch.}
\label{figeqstat}
\end{figure}

\begin{figure}[p]
\resizebox{3.0in}{!}{\includegraphics{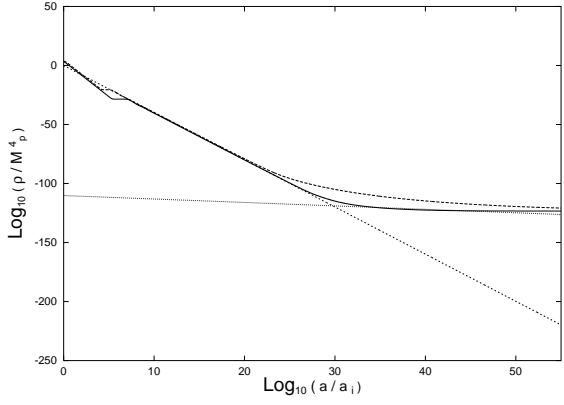}}
\caption{The evolution of energy density in the field is shown for two values of $g_c=0.9$(solid line) and  $g_c=0.07$(dotted line) for
the
model described (\ref{pottw+}) with $\lambda=0.52$. The radiation energy density is depicted by the dashed line. The
second dashed line below shows the(non-coupled) inflationary behavior of $\rho_{\phi}$($ \rho_{\phi} \propto a^{-\lambda^2}$). The decay of $\rho_\phi$ slows down earlier for smaller value of the coupling constant $g_c$ and
 approaches the inflationary attractor behavior after overtaking   the background(radiation) energy density. After staying with the attractor value 
for quite some time, $\rho_{\phi}$ becomes constant.}
\label{figdeng1g2}
\end{figure}

\begin{figure}[p]
\resizebox{3.0in}{!}{\includegraphics{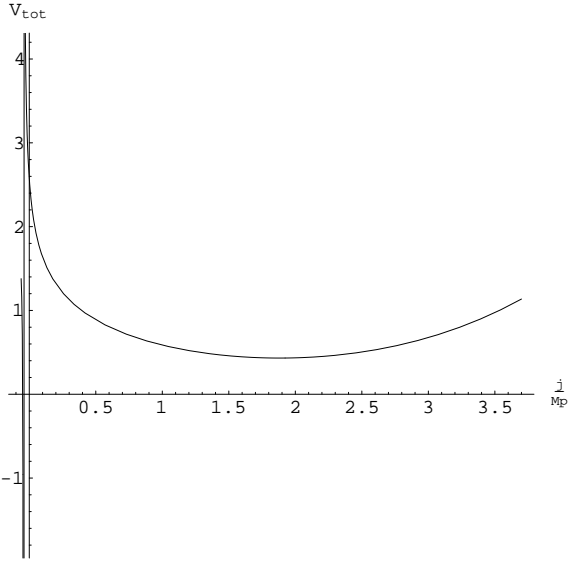}}
\caption{The effective potential in equation (\ref{veffpot}) for $\lambda =0.52, g_c=5.9$.  
This  effective potential becomes repulsive near the origin
when $(\phi_{cr}/M_p)=1.88$ followed by an infinite barrier at $(\phi^b/M_p)=-0.042$ which shields the singularity situated at $(\phi^s/M_p)=-0.084$.}
\label{figeffpot}
\end{figure}
\begin{figure}[p]
\resizebox{3.0in}{!}{\includegraphics{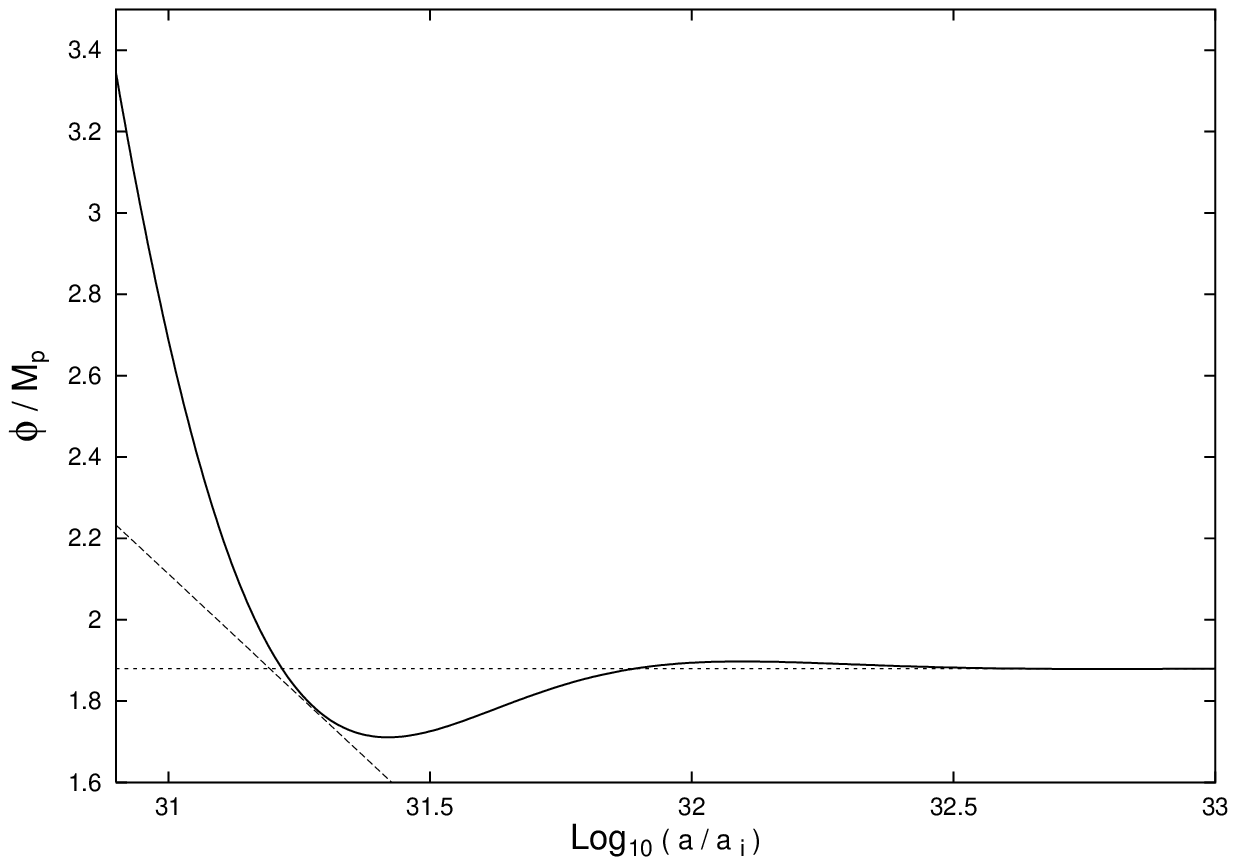}}
\caption{Late time evolution of the scalar field (solid line). After the current
epoch, the scalar field passes across the critical point, moves towards the barrier and gets reflected
in the effective potential shown in the previous figure. It makes few oscillations
and settles at the critical point ($ \phi_{cr}/M_p) = 1.88$.
The dashed line depicts the inflationary attractor behaviour of the scalar field in case of the exponential potential (\ref{pottw+})
with $\lambda=0.52$ ($\phi/M_p=-\lambda \ln(a/a_i)+constant$) and the dotted line corresponds to $(\phi_{cr}/M_p)=1.88$.}
\label{figft}
\end{figure}

\section{QUINTESSENCE WITH EXPONENTIAL POTENTIAL IN THE PRESENCE OF THE COUPLING TO THE TRACE $T$ }
We shall now consider a model with
\begin{equation}
V(\phi)=V_0 \exp({\lambda{ \phi \over M_p}}),\qquad (\lambda < \sqrt{2})
\label{pottw+}
\end{equation}
in which the field  evolves towards the
origin rather than away from the origin.
 In order to investigate the dynamics described by  (\ref{evoleqn}) and (\ref{friedeqn}), it would be convenient to cast these equations as a system of first order equations
\begin{equation}
Y_1'=Y_2
\label{eqneight}
\end{equation}
\begin{eqnarray}
Y'_2&=& -3Y_2+{1 \over H(Y_1, Y_2)}\Big[{g_c \over {(1+2g_cY_1)}}
Y_2^2\nonumber \\
 &+&g_c{{(1+2g_cY_1)} \over {(1+4g_cY_1)^2}} \tilde{V}(Y_1) -{d\tilde{V}(Y_1) \over dY_1}{{(1+2g_cY_1)} \over{(1+4g_cY_1)}} \Big]\nonumber\\
 &+& {\rho^i_m \over a^3} g_c {(1+2g_cY_1) \over {(1+g_cY_1)^2}}
\end{eqnarray}
where
 \begin{equation}
 Y_1={\phi \over M_p},\quad Y_2={\dot{\phi} \over M_p^2},\quad \tilde{V}={ V(Y_1) \over M_p^4} 
 \end{equation}
and prime denotes the derivative with respect to the variable $N=\ln(a)$. The function
$ H(Y_1, Y_2)$ is given:
\begin{equation}
 H(Y_1, Y_2)=\sqrt{{1 \over 3}\left[{Y_2^2 \over {2(1+2g_cY_1)}}+{\tilde{V}(Y_1) \over {(1+4g_cY_1)}}+{\rho_b \over M_p^4}  \right]}
\end{equation}
where $\rho_b =\rho^i_r e^{-4N}+\rho^i_m e^{-3N}/(1+g_cY_1)$.
The initial conditions will be set at Planck time
with $
a_{i}=1$. 
(The scale factor  today would nearly be  $10^{31}$.)
Further, for the background energy density,
 $\rho_r=\rho^i_r/a^4$, we make the choice,
$\rho^i_r=M_p^4$
 based on the point of view that, at the 
radiation dominated phase near the Planck's epoch, the energy density of radiation
was of the order of Planck energy as it was the only scale available at
that time. The initial values for $\phi$ and $\dot\phi$ as well as the the parameters in the potential 
are chosen so as to ensure a viable cosmological model. We take the initial values  $Y_1=550$;
$Y_2=0$ and $\lambda=0.52$, $ g_c=5.9$. The value of $V_0$ is chosen so that initially, $\rho^i_{\phi}/\rho^i_R \simeq 10^2$, 
Of these choices, $\lambda$ and $ g_c$ are important for obtaining a viable model;  $Y_1$ is arbitrary 
except for the condition  $Y_1 \gg 1$. 
The choices made for $Y_2$ and $ V_0 $ are merely done for the sake of convenience; different choices 
will lead to same qualitative behaviour. 
Some of these choices could be construed as fine tuning in the model. But it is no worse than the 
fine tuning in any other scalar field model.

The model described by (\ref{pottw+}) supports an eternal inflation in absence of the coupling to the trace $T$ . The forcing term induced by the back
reaction in the field evolution equation acts against the Hubble damping and hastens the evolution of field energy density $\rho_{\phi}$.
We have plotted the ratio of kinetic to potential energy of the field in figure \ref{figkp+}. It is seen that due to coupling to the trace $T$ , the kinetic energy
 acquires a large value making the potential term unimportant
and pushing the field into kinetic regime. As a result, the field energy density crosses the background energy density and continues
evolving  as $(1/a^6)$ for quite some time  (see figure \ref{figden1}). After the crossover, the dominant contribution to the Friedman equation (\ref{friedeqn}) comes from the
background energy density  $\rho_b$. This in turn leads to the increase in damping and consequently slowing down of the decay of $\rho_\phi$
allowing the ratio $(\rho_{\phi}/\rho_{b})$ to grow. The energy density gets locked to a constant value in this process and acts like cosmological constant thereby making $\Omega_{\phi}$ to grow. The forcing term containing $\dot{\phi}^2$ in the field evolution
equation is ineffective in this regime. 

A remarkable 
thing happens when the field starts rolling again at the end of locking regime. As discussed above, the ratio of kinetic to potential
energy of the field ($K_e/P_e$)  provides a yardstick to determine the energy scaling.
 At the end of locking period, the field again starts rolling down the hill.
Though back reaction  again builds up, the kinetic to potential energy ratio is now much smaller  than it was in the beginning
when  $\rho_{\phi}$ was dominant. 
The ratio  first increases, then fluctuates near
its background value, finally heading towards zero  (figure \ref{figkp1+}). As a result, $\rho_{\phi}$ first
(nearly) tracks the $\rho_b$, gradually approaches it and ultimately  overtakes it. This allows the scalar field
to dominate and gives rise to  accelerated expansion at late times.
By suitably tuning the parameters of the model, it is possible to get
$\Omega_{\phi}=0.7$ today  (figure \ref{figomega}). In figure \ref{figoustat}, we have displayed the equation of
state  parameter $w_{\phi}$  to demonstrate
the convergence of different initial conditions to the inflationary
attractor. Figure (\ref{figeqstat}) shows that the $w_{\phi}$ mimics de-Sitter like behavior shortly after the current epoch.

The role
of $\dot{\phi}^2$ term on the right-hand-side of evolution equation (\ref{evoleqn}) is very important in the beginning. The size of the undershoot crucially depends upon the coupling constant $g_s$.
Larger value of $g_s$ leads to deeper undershoot because the ratio $K_e/P_e$ is sensitive to coupling initially. After the locking
period this ratio of the order of few as background (radiation) dominates over the field this time. During evolution, the scalar
field moves from larger values towards the origin. As a result the influence of the back reaction in the Friedman equation
(\ref{friedeqn}) decreases
 and the field energy density gradually moves towards the background, overtakes it and joins with the (non-coupled) inflationary attractor(This may be contrasted with the behaviour in the 
model described by (\ref{exppot}), in which the scalar field evolves from origin to larger values. In this case the contribution of
field energy density in the Friedman equation gets gradually suppressed allowing the background to 
dominate for ever. In this model
$\rho_{\phi}$ tracks the background and never comes to dominate it(see figures, \ref{figone},\ref{figkp1} and \ref{figtwo}). 
This
development is displayed in figure (\ref{figdeng1g2}) which clearly shows: (i) deeper undershoot for larger coupling and (ii) joining of the inflationary track
 earlier for smaller value of $g_s$. Figure \ref{figdeng1g2} also shows that the field continues to stay in the usual
(non-coupled) attractor phase for quite some time till the  potential terms induced by the coupling in field evolution equation become important. This happens at
very late times and the influence of these terms will be discussed below. As for the matter term, numerical analysis shows that they do not influence the dynamics significantly. Thus coupling plays an important role in the intermediate regime(which also maifests in figures \ref{figden1},\ref{figoustat} and \ref{figeqstat}) thereby allowing to produce current observed
accelerated expansion with necessary tuning of $g_c$ along with $\alpha$.

Let us briefly summarize the role of the coupling to the trace $T$ .The
exponential potential under consideration is shallow and supports an
eternal inflation in the standard scenario. The introduction of the
coupling to the trace $T$  hastens the decay of $\rho_\phi$, pushing the scalar field into
kinetic regime. After the locking period, its role gradually
diminishes making it possible for the scalar field, at late times, to
join back the inflationary track it was detracted from.
At this stage, the scalar field is  approaching the origin and we would expect the equation
 of state $w_{\phi}$  to settle in its attractor value. However, the equation of state becomes equal to 
 $(-1)$ and stays there ---
which is evident from figure \ref{figeqstat}. The same effect is reflected in the behavior of $\rho_{\phi}$ shown in figure 
\ref{figdeng1g2}, where the field energy
density becomes constant after staying in its attractor value for some time. This seems to be a general feature of the model under
consideration and requires explanation.
 
The coupling to the trace $T$  modifies the field evolution through  three extra terms on the right hand side  of (16). The first  term contains
$\dot{\phi}^2$ and plays important at the early epochs and during tracking as discussed above. The second term contains the potential
which contributes in the potential dominated regimes: during the turn around of $\rho_\phi$ and
 when the field is rolling
slowly.  We will show that this term is responsible for the cosmological constant like behavior of the scalar field at very late times
, i.e., $a\to \infty$.
In order to see this, we will cast (\ref{eqneight})  in the form
\begin{equation}
\dot{Y_2}+{ \dot{a} \over a}{Y_2}=g_c{{Y^2_2} \over {1+2g_c Y_1}}-{d \over dY_1}V_{tot}(Y_1)
\label{seventeen}
\end{equation}
where $V_{\rm tot}(Y_1)$ is the effective potential given by
\begin{equation}
V_{\rm tot}(Y_1)=\int{ \left[ \lambda{{(1+2g_cY_1)} \over {(1+4 g_c Y_1)}} 
 -g_c {{(1+2g_cY_1)} \over {(1+4 g_c Y_1)^2}}\right]\tilde{V}(Y_1)} dY_1 
 \label{veffpot}
\end{equation}
where $\tilde{V}=(V_0 / M^4_p) e^{\lambda Y_1}$.
(The matter term is sub-dominant in this limit and is hence ignored.)
In the model described by (\ref{pottw+}), the scalar field is evolving from large values towards the origin and can take negative values. But when $Y_1\to -(1 /2 g_c)$ 
 the system will become singular because of the divergence of the first term in (\ref{seventeen}). However, the second term in the expression of $V_{tot}(Y_1)$ does not let it happen. In fact, this term sets an infinite potential barrier at $Y_1=-(1 /4g_c)$
and does not allow the field to reach the singularity. The potential which was attractive away from the origin becomes now repulsive
followed by the infinite barrier on its left (see  figure \ref{figeffpot}). This happens when
\begin{equation}
Y^{cr}_1={{4g_c-\lambda} \over {4 \lambda g_c}}
\end{equation}
At this stage, the field $\phi$ is rolling slowly. It crosses the critical point and moves towards the barrier but due to 
insufficient kinetic energy gets  reflected back by the barrier. After making few such oscillations, the field settles at
the critical point permanently and mimics the cosmological constant. For the model described by (\ref{pottw+}) with $ \lambda=0.52$ and
$g_c=5.9$, $Y^{cr}_1 \equiv \phi^{cr}/M_p \simeq 1.88$ which is confirmed by numerical integration (see, figure
\ref{figft}).

\begin{figure}[p]
\resizebox{3.0in}{!}{\includegraphics{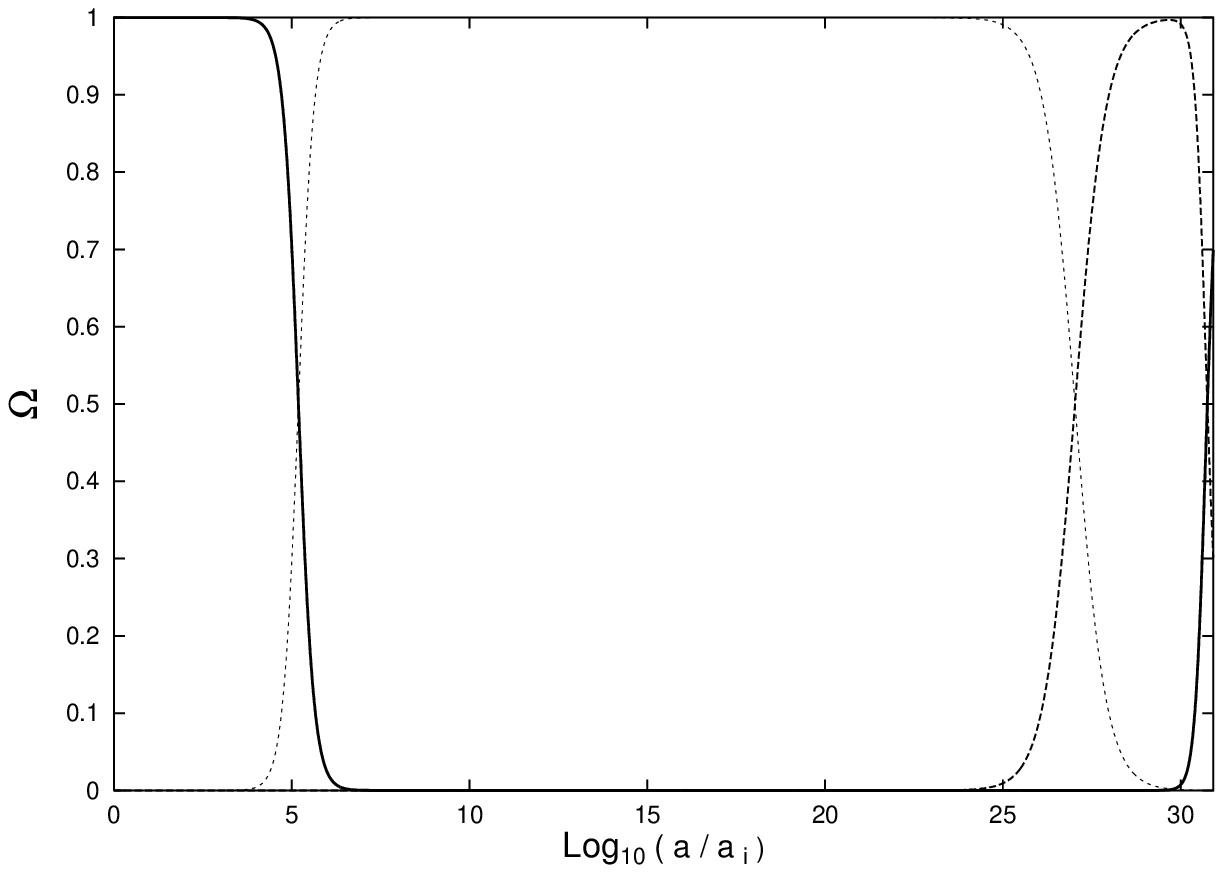}}
\caption{The dimensionless density parameter is plotted as a function of the scale factor for:
(i) scalar field (solid line)
with a potential $V = V_0 \sinh^{2p}(\lambda \phi/2M_p)$,
  radiation ( dotted line ) and matter (dashed line).
The late time behavior of the scalar field is shown to lead  to the present values of   $\Omega_{\phi}=0.7$ and
 $\Omega_m=0.3 $}
\label{figomegainf}
\end{figure}

\section{ MODELS OF QUINTESSENCE  FREE FROM FUTURE EVENT HORIZON PROBLEM}
\begin{figure}[p]
\resizebox{3.0in}{!}{\includegraphics{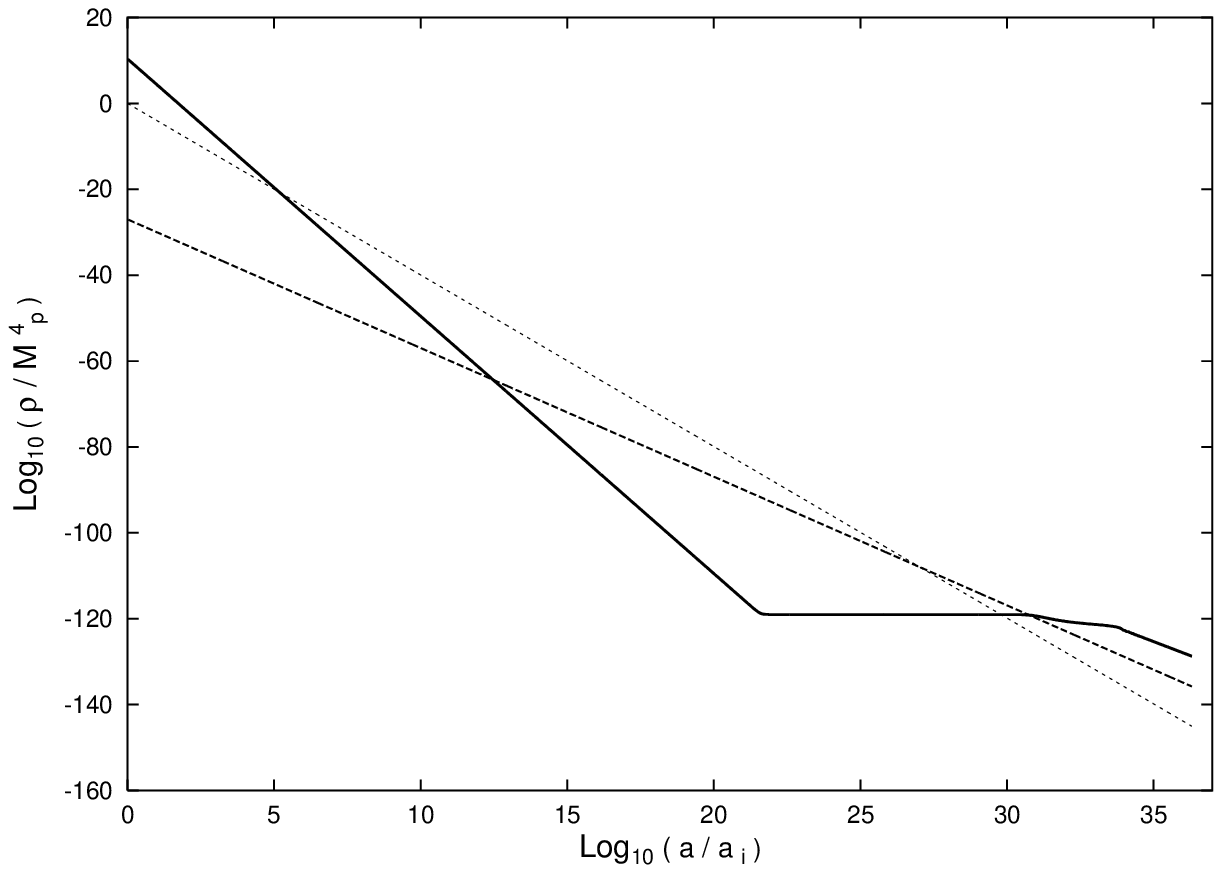}}
\caption{Evolution of energy density is shown for scalar field (solid line) 
 radiation (dotted) and matter (dashed) for a potential $V\propto \sinh^{2p}(\lambda\phi/2M_p)$. The 
        energy density $\rho_{\phi}$ drops below $\rho_{\rm bg}$, turns around catching up with it in the matter dominated
        era. The scalar field dominates at late times and derives the
        current cosmic acceleration. The universe continues in this
        phase for  some time till field oscillations ( near the
        origin $\phi=0$ ) build up in the system leading to average
        equation of state $<w_{\phi}>=0$. The field behaves as cold
        dark matter thereafter.}
\label{figdeninf}
\end{figure}

In the model described above, like in any generic model of quintessence, universe will be accelerating forever leading to future event horizon --- which could pose a problem
in the string theoretical context. It is, however, possible to tackle this problem  by using any potential which is exponential for large $\phi$ and is a power law near the origin: That is, we require $V(\phi)\to \exp(\lambda\phi/M_P)$ for large $\phi$ and 
$V(\phi)\to (\lambda\phi/M_P)^{2p}$ for $\phi\to 0$. 
The power law behaviour near the origin will lead  to oscillations
when the field approaches $\phi \simeq 0$. The mean equation of state of the scalar field is 
then given by \cite{turner}
\begin{equation}
< w_{\phi} > \simeq \left <{{ {\dot{\phi}^2 \over 2}-V(\phi)} \over {{\dot{\phi}^2 \over 2}+V(\phi)}} \right >={{p-1} \over{p+1}}
\end{equation}
As a result the scalar field energy density and scale factor have the following behavior
$$\rho_{\phi} \propto  a^{-3(1+< w >)},~~~~~~~~a \propto   t^{{ 2 \over 3}(1 + < w >  )^{-1}} $$
The scalar field behaves like pressure less dust for $p=1$. By suitably adjusting the parameters in the model one could  ensure that the oscillations occur in the 
 future, i.e., well after  the present epoch when $\Omega_{\phi}$ has reached value equal to 0.7. One can avoid the future event horizon in such a model,  irrespective of the explicit form of $V(\phi)$
which is used, as long as it interpolates between an exponential and power law behaviour.
One of the many possible choices of interpolation is a potential of the form $V(\phi)=V_0
\sinh^{2p}(\lambda \phi/2M_p)$, for which 
we have evolved the field
equations numerically. The results for a particular choice of parameters are shown in 
figures (\ref{figomegainf}) and (\ref{figdeninf}).

We  summarize below the chronology of main events in any such model with a shallow potential which behaves like a power law near the origin:
(i) The scalar field starts rolling slowly in the shallow potential. The coupling to the trace $T$  almost instantaneously kills
any inflationary behavior  and makes  the scalar field to enter into the kinetic regime. The scalar field
continues to stay in the kinetic regime (with $\rho_\phi\propto a^{-6}$) and soon $\rho_\phi$ drops below  the $\rho_{b}$. 
(ii) Once this happens, the dominant contribution to  the Friedmann equation comes from the background. This, in turn, allows $\rho_{\phi}$ to turn towards $\rho_{b}$ leading to a locking regime during which  the value of $\rho_{\phi}$ stabilizes and remains relatively unchanged for a considerable length of time and the scalar field behaves like cosmological constant.
(iii) The field  starts rolling down the potential again and  the back reaction builds up such 
that the kinetic energy of the field
exceeds its potential energy by a factor of few in the beginning. However, as the field rolls 
towards smaller and smaller values, the
role of back reaction gradually diminishes and the potential energy starts catching up again with the kinetic energy. 
As a result $\rho_{\phi}$ slowly approaches the back ground energy density and finally overtakes it.
 The scalar field is now back on the original inflationary
track. This phase  is interpreted as the late time accelerated expansion of the universe,
leading to present day
value of $\Omega_{\phi}=0.7$. 
(iv) After the current epoch, the universe continues in the phase of cosmic acceleration for quite some time till
 the scalar field approaches $\phi \simeq 0 $ and starts oscillating about it giving rise to $<w_{\phi}>=0$ and
 turning the scalar field to behave like cold dark matter (see figure \ref{figdeninf}). This is expected to happen in future and  the model is
 free from the event horizon problem at late times. (There are other possible solutions to this problem; see,
 for example, 
 \cite{bjp, shtanov}).

\begin{figure}[p]
\resizebox{3.0in}{!}{\includegraphics{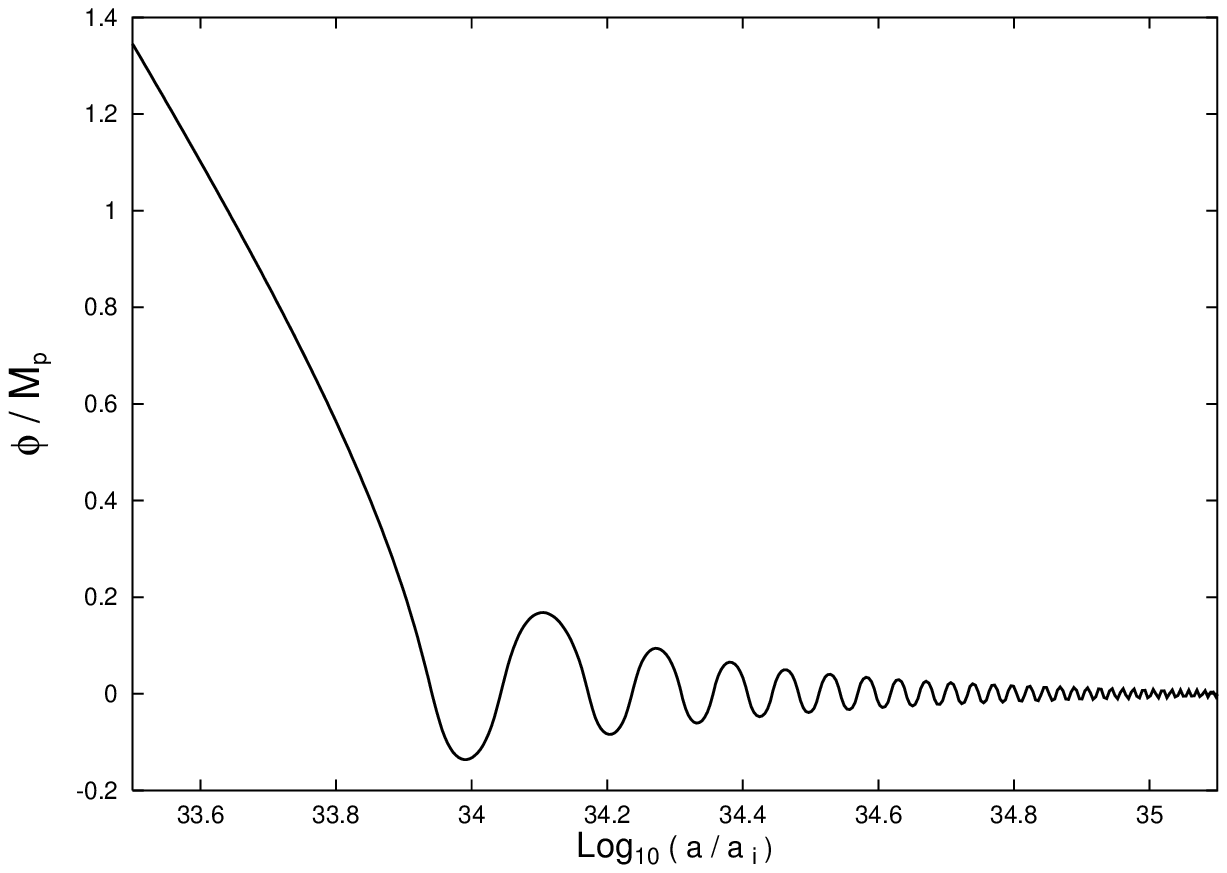}}
\caption{Evolution of the scalar field is shown for the model described in figure \ref{figdeninf} 
in the asymptotic future. (The current epoch is $a\approx 10^{31}$.)
After
driving the accelerated expansion for quite some time, the field enters into oscillatory phase near the origin with the
average equation of state $<w_{\phi}>=0$.}
\label{figfield}
\end{figure}

\section{CONCLUSIONS}
We have examined the dynamics of a scalar field  which couples to the trace of all the field including itself in FRW background.
The coupling to the trace $T$  of the field modifies the energy momentum tensor and induces a forcing term in the field evolution equation. The
forcing term acts against the Hubble damping changing the evolution  of field energy density significantly. In case of a steep potential,
it amounts to increasing  the steepness of the potential. In case of the shallow potentials in which the field rolls
slowly and mimics cosmological constant, the coupling to the trace $T$    plays an important role. It forces the scalar field into kinetic
regime allowing the  radiation domination to commence. We have discussed a model in which the effect of the back reaction gradually diminishes allowing the scalar field to catch up with the inflationary attractor regime at late times, thereby accounting for the observed 
cosmic acceleration with $\Omega_{\phi}=0.7$. We have also presented a  variant of the model in which the scalar field plays the role of
quintessence at the current epoch and mimics  dust like dark matter in future which eliminates 
the future event horizon. The general features of the model are shown to be independent of the initial 
conditions. However,  fine tuning of some of the parameters  is necessary to achieve viable
evolution.

\appendix

\section{ Action for scalar field coupled to the trace of stress tensor}

We summarize here the derivation of the self-consistent Lagrangian for a scalar field coupled to the trace of the energy momentum tensor of all matter, originally given in \cite{tpstp88}, for the sake of completeness.
  
  Consider a system consisting of the gravitational field $g_{ab}$, radiation fields, and a  scalar field 
  $\phi$ which couples to the trace of the energy-momentum tensor of all fields, including its own. 
  The {\em zeroth order} action for this system is given by
  \begin{equation}
  A^{(0)} = A_{\rm grav} + A_\phi^{(0)} + A_{\rm int}^{(0)} + A_{\rm radn}
  \end{equation}
  where 
  \begin{equation}
  A_{\rm grav} = (16 \pi G )^{-1} \int R \sqrt{-g} \, d^4x - \int \Lambda \sqrt{-g} \, d^4x,
  \end{equation}
  \begin{equation}
  A_\phi^{(0)} = \frac{1}{2} \int \phi^i\phi_i\sqrt{-g} \, d^4x; \quad A_{\rm int}^{(0)} = \eta \int Tf(\phi/\phi_0) 
  \sqrt{-g} \, d^4x
  \end{equation}
 Here, we have explicitly included the cosmological constant term and $\eta $ is a dimensionless number which 
 `switches on' the interaction. In the zeroth order action, $T$ represents the trace of all fields other 
 than $\phi$. Since the radiation field is traceless, the only zeroth-order contribution to $T$ comes from the $\Lambda$ term, so that we have $T= 4 \Lambda$. The coupling to the trace is through a function $f$ of the scalar
 field, and one can consider various possibilities for this function. The constant $\phi_0$ converts $\phi$
 to a dimensionless variable, and is introduced for dimensional convenience.

 Since the stress-tensor of the scalar field has a non-zero trace,
  we must add to $T$ the contribution
  $T_\phi = - \phi^l\phi_l$  of the scalar field. 
  If we now add $T_\phi$ to $T$ in the interaction term $A_{\rm int}^{(0)}$ further modifies $T^{ik}_\phi$.
  This again changes $T_\phi$. Thus to arrive at the correct action an infinite iteration will have to be performed and the complete action  can be obtained by summing up all the terms (see \cite{jvntp}). The full action  can be found, more simply, by a  consistency
  argument.
  
  Since the effect of the iteration is to modify the expression for $A_\phi$ and $A_\Lambda$,  we consider
  the following ansatz for the full action:
  \begin{eqnarray}
  A &=& \frac{1}{16 \pi G} \int  R \sqrt{-g} \, d^4x - \int \alpha(\phi) \Lambda  \sqrt{-g} \, d^4x
  \nonumber \\
  && \quad +\frac{1}{2}
  \int \beta(\phi) \phi^i\phi_i  \sqrt{-g} \, d^4x  + A_{\rm rad}
  \end{eqnarray}
  Here $\alpha(\phi)$ and $\beta(\phi)$ are functions of $\phi$ to be determined by the 
  consistency requirement  that they represent the result of the iteration of the interaction term. (Since radiation makes no contribution to $T$, we expect  $A_{\rm rad}$ to remain unchanged.)
  The energy-momentum tensor for $\phi$ and $\Lambda$ is now given by
  \begin{equation}
  T^{ik} = \alpha(\phi) \Lambda g^{ik} + \beta(\phi) \left[ \phi^i\phi_k - \frac{1}{2} g^{ik} \phi^\alpha\phi_\alpha\right]
  \end{equation}
  so that the total trace is $T_{\rm tot}=4\alpha(\phi)\Lambda - \beta(\phi) \phi^i\phi_i$. The functions
  $\alpha(\phi) $ and $\beta(\phi)$ can now be  determined by the consistency requirement
  \begin{widetext}
  \begin{eqnarray}
  && -\int \alpha(\phi) \Lambda  \sqrt{-g} \, d^4x  +\frac{1}{2}
  \int \beta(\phi) \phi^i\phi_i  \sqrt{-g} \, d^4x \nonumber \\
  && \quad = -\int \Lambda \sqrt{-g} \, d^4x +  \frac{1}{2}
  \int \phi^i\phi_i  \sqrt{-g} \, d^4x +\eta \int T_{\rm tot} f(\phi/\phi_0) \sqrt{-g} \, d^4x
  \end{eqnarray}
   \end{widetext}
  Using $T_{\rm tot}$ and comparing terms in the above equation we find that
  \begin{equation}
  \alpha(\phi) = [ 1+ 4 \eta f]^{-1}, \quad \beta(\phi) = [ 1+ 2 \eta f]^{-1}
  \end{equation}
  Thus the complete action can be written as
  \begin{eqnarray}
  A &=& \frac{1}{16 \pi G} \int  R \sqrt{-g} \, d^4x - \int \frac{ \Lambda}{1+4\eta f}  \sqrt{-g} \, d^4x
  \nonumber \\
  && \quad+ \frac{1}{2}
  \int \frac{ \phi^i\phi_i }{1+2\eta f} \sqrt{-g} \, d^4x  + A_{\rm rad}
  \label{completeaction}
  \end{eqnarray}
  (The same action would have been obtained if one uses the iteration procedure.) 
  It is obvious that the method works for any source for which the trace of stress tensor, $T$,
  is proportional to the Lagrangian $L$. If $T=\mu L$, then the coupling replaces $L$ by 
  $L(1-\mu\eta f)^{-1}$. [For the kinetic energy term $L_\phi =
  (1/2)\phi_a\phi^a$, we have, $T_\phi = -\phi^a\phi_a = (-2) L_\phi$ 
  and for the cosmological constant term $T_\Lambda = 4\Lambda = (-4)L_\Lambda$.
  These lead to the factor $(1+2\eta f)^{-1}$ and $(1+4\eta f)^{-1}$ in (\ref{completeaction}).] In general,
  if the action is a homogeneous function of degree $D$ in $g^{ab}$, the $T=(2D) L$ and the coupling 
  factor is $(1-2D\eta f)^{-1}$. For a system of dust like particles 
  with the action for each particle being the integral of $ds= \sqrt{g_{ab} dx^a dx^b}$, we have
  $D=(-1/2)$ and the coupling factor becomes $(1+\eta f)^{-1}$ (see \cite{jvntp}).
  
  In the presence of cosmological constant and sources which are conformally invariant, the action in (\ref{completeaction}) leads to the following field equations,
  \begin{widetext}
  \begin{equation}
  R_{ik} - \frac{1}{2} g_{ik} R = -8 \pi G \left[ \beta(\phi) \left( \phi^i\phi^k - \frac{1}{2} g^{ik} \phi^\alpha\phi_\alpha \right) + \frac{\Lambda}{8\pi G} \alpha(\phi) g_{ik} +T_{ik}^{\rm traceless}\right]
  \end{equation}
  \end{widetext}
  \begin{equation}
  \Box \phi +\frac{1}{2} \frac{\beta'(\phi)}{\beta(\phi)} \phi^i\phi_i + \frac{\Lambda}{8\pi G} \frac{\alpha'(\phi)}{\beta(\phi)} =0
  \end{equation} 
  Here, $\Box$ stands for a covariant d'Lambertian, $T_{ik}^{\rm traceless}$ is the 
  stress tensor of all fields with traceless stress tensor and a prime denotes differentiation with respect 
  to $\phi$.
  
  In the cosmological context, this reduces to
  \begin{equation}
  \ddot \phi +\frac{3\dot a}{a} \dot \phi = \eta \dot \phi^2 \frac{f'}{1+2\eta f} + \eta \frac{\Lambda}{2\pi G} \frac{f'(1+2\eta f)}{(1+4\eta f)^2}
  \label{aeleven}
  \end{equation}
  \begin{equation}
  \frac{\dot a^2 +k}{a^2} = \frac{8\pi G}{3} \left[ \frac{1}{2} \frac{\dot \phi^2}{1+2\eta f} + \frac{\Lambda}{8\pi G}
  \frac{1}{(1+4\eta f)} + {\rho_b(a)}\right]
  \label{atwelve}
  \end{equation}
where $\rho_b(a)=\rho^i_R/a^4$.
  It is obvious that the effective cosmological constant can decrease if $f$ increases in an expanding 
  universe. The result can be easily generalized for a scalar field with a potential by
  replacing $\Lambda$ by $V(\phi)$ in the action.
  Similarly, the presence of dust like matter adds the term $(\rho_i/a^3)(1+\eta f)^{-1}$ to the 
  source term in (\ref{atwelve}) and a corresponding derivative term in (\ref{aeleven}).

\begin{figure}[p]
\resizebox{3.0in}{!}{\includegraphics{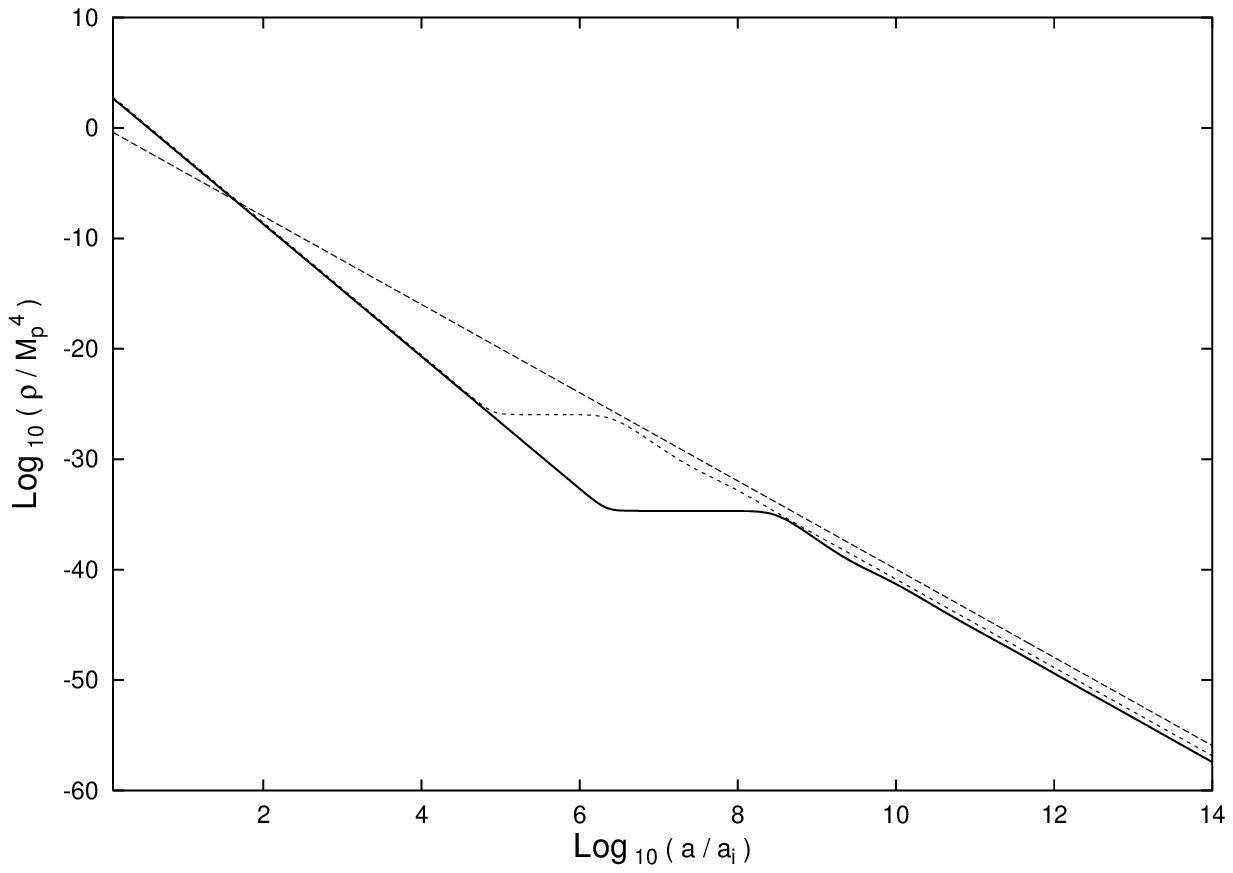}}
\caption{ The evolution of scalar field density corresponding to $g_c=0.05$ (solid line) and $g_c=0$ (dotted line)
is shown for the potential (\ref{exppot}) with $\lambda=6$. The radiation energy density is shown by the dashed line. The  pattern of tracking shows that the effect of forcing
terms amounts to  increase the steepness of the potential.}
\label{figone}
\end{figure}
  
\begin{figure}[p]
\resizebox{3.0in}{!}{\includegraphics{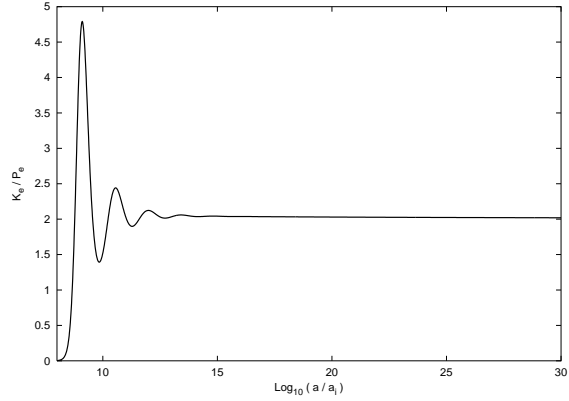}}
\caption{Evolution of the kinetic to potential energy ratio of scalar field for the potential (\ref{exppot}) with $\lambda=1.2$ and $g_c=1$ starting
from the locking regime onwards. At the end of locking period, the ratio fluctuates about its background (radiation in this case) value ($ K_e/P_e=2 $)
and ultimately approaches it.}
\label{figkp1}
\end{figure}

\begin{figure}[p]
\resizebox{3.0in}{!}{\includegraphics{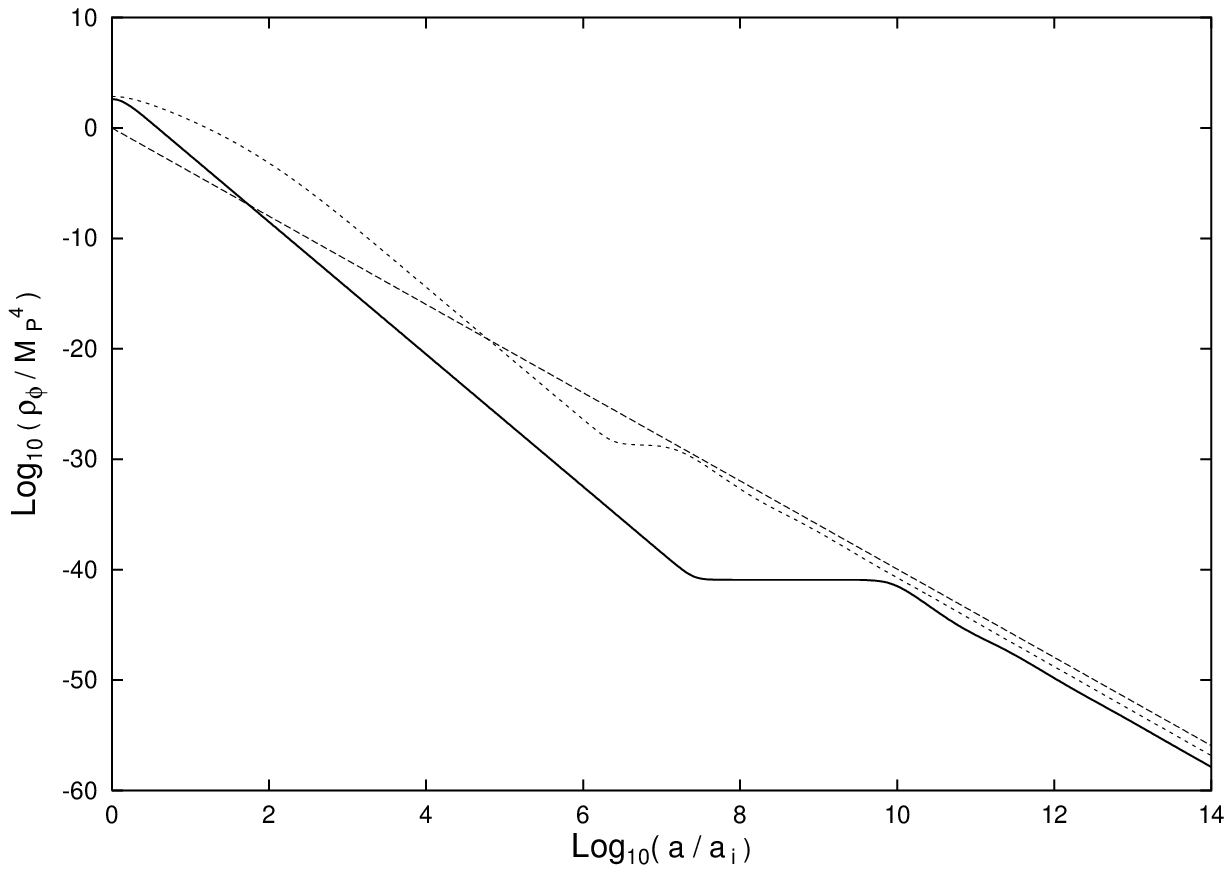}}
\caption{The energy density of scalar field $\rho_{\phi}$ is plotted for $g_c=1$ (solid line) and $g_c=0.1$ (dotted line) for shallow exponential
potential (\ref{exppot}) with $\lambda=1.2$. The dashed line shows radiation energy density. The effect of the forcing term in the evolution equations is shown to accelerate the scaling of $\rho_{\phi}$. The stronger coupling is seen to hasten the commencement of kinetic
regime. The scalar field ultimately tracks the back ground and gets locked up there for ever.}
\label{figtwo}
\end{figure}

\section{COSMOLOGY WITH $V\propto \exp(-\lambda\phi/M_P)$ AND COUPLING TO T}

In this case, the field evolves to large $\phi$ at late times making the matter term $a^{-3}[1+\eta f]^{-1}$
sub-dominant to the radiation term $a^{-4}$. We will, therefore, ignore the matter term for simplicity.
We first discuss the  possibility when $ \lambda>\sqrt{n}$. In this case the potential is steep leading to   $\rho_{\phi}
\propto 1/a^6$. This makes $\rho_{\phi}$ sub-dominant relative to $\rho_b$ and the motion will become strongly damped allowing
the kinetic energy to decrease. This, in turn, allows  $\rho_{\phi}$  to catch up  with the background
and stay there for ever till the attractor regime is reached. The role of coupling to $T$   will be to increase
the kinetic energy and keep  $\rho_{\phi}$ in the kinetic regime for longer. 
The  $\rho_{\phi}$ will catch up again with the background,
 though  it takes longer to reach the attractor (scaling) solution with the duration depends upon $g_c$. Therefore, in this case,
the role of the forcing term is equivalent to increasing the steepness of the potential given by $\lambda$ (see figure \ref{figone}). Thus nothing new or interesting happens
in this case .

The case with $\lambda < \sqrt n$ ( $n=4 $ in our case) is more interesting. We shall be working with  
$\lambda< \sqrt 2$
in which case the exponential potential supports a never ending inflation. The forcing term which acts against damping.  
 Due to coupling to the trace $T$  the kinetic energy  acquires a large value making the potential term unimportant
and pushing the field into kinetic regime. 
Depending upon the value of coupling $g_c$
it may take longer or shorter to force $\rho_{\phi}$ to come  into   kinetic regime which  then crosses the background
energy density, (see figure \ref{figtwo}). The extent of the overshoot depends upon the coupling constant $g_c$. The evolution
of $\rho_{\phi}$ in this model  proceeds more ore less in the similar fashion as in case of the model described by (\ref{pottw+}) till the turn around. 
 At the end of locking period, the field again starts rolling down the  potential.
Though back reaction again builds up, the kinetic to potential energy ratio is now much smaller  than it was in the beginning
when  $\rho_{\phi}$ was dominant (figure \ref{figkp1}). The field does not
join the inflationary attractor in the present case. Because as it runs down the hill away from the origin to larger
and larger
values, the contribution of $V(\phi)$ and $\dot{\phi}^2/2$ to the right hand side  of Friedmann equation (\ref{friedeqn}) gets suppressed systematically
and   $\rho_{\phi}$ remains sub-dominant to the background (radiation). The ratio $(K_e/P_e)$ settles at value of 2 ( $w_{\phi} =1/3$) and
the scalar field tracks the background and gets locked there for ever.

\begin{acknowledgements}
One of us (MS) is  thankful to V. Sahni and T. Qureshi for useful discussions. 
 
\end{acknowledgements}

\end{document}